\documentclass{ws-procs9x6}

\usepackage{graphicx,wrapfig,amssymb}
\usepackage[T1]{fontenc}
\usepackage{ae,aecompl}

\newcommand{\be}{\begin{equation}}
\newcommand{\ee}{\end{equation}}
\newcommand{\ba}{\begin{eqnarray}}
\newcommand{\ea}{\end{eqnarray}}
\newcommand{\la}{\langle}
\newcommand{\ra}{\rangle}
\newcommand{\di}{ {\rm d} }

\begin{document}

\title{Sivers effect in semi-inclusive deeply inelastic scattering
and Drell-Yan}

\author{J.~C.~Collins$^{1,3}$, A.~V.~Efremov$^2$, K.~Goeke$^3$, \\
        M.~Grosse~Perdekamp$^{4,5}$, S.~Menzel$^3$, B.~Meredith$^4$, \\
	A.~Metz$^3$, P.~Schweitzer$^3$}
\address{
$^1$ Penn State University, 104 Davey Lab, University Park PA 16802, U.S.A.\\
$^2$ Joint Institute for Nuclear Research, Dubna, 141980 Russia\\
$^3$ Institut f{\"u}r Theoretische Physik II, Ruhr-Universit{\"a}t Bochum, Germany\\
$^4$ University of Illinois, Department of Physics, Urbana, IL 61801, U.S.A.\\
$^5$ RIKEN BNL Research Center, Upton, New York 11973, U.S.A.
\vspace{-1cm}}
\maketitle
\abstracts{
  The Sivers function is extracted from HERMES data on single spin asymmetries 
  in semi-inclusive deeply inelastic scattering. The result is used for making
  predictions for the Sivers effect in the Drell-Yan process.
  \vspace{-1cm}}

%===================  SECTION 1: INTRODUCTION ========================
\section{Introduction}
\label{Sec-1:introduction}

The recent HERMES and COMPASS 
data\cite{HERMES-new,Airapetian:2004tw,Alexakhin:2005iw,Diefenthaler:2005gx}
on transverse target single spin asymmetries (SSA) in semi-inclusive deeply 
inelastic scattering (SIDIS) can be described\cite{Boer:1997nt}
--- on the basis of generalized factorization 
theorems\cite{Collins:1981uk,Ji:2004wu,Collins:2004nx} --- in terms of the 
Sivers\cite{Sivers:1989cc,Brodsky:2002cx,Collins:2002kn,Belitsky:2002sm} 
and Collins\cite{Collins:1992kk} effects. These effects may also contribute to 
SSA in hadron-hadron-collisions\cite{Adams:1991rw} and longitudinal SSA in 
SIDIS\cite{Avakian:1999rr,Airapetian:1999tv,Airapetian:2002mf,Avakian:2003pk},
but the status of factorization for these SSA is not clear.
All that is clear is that the longitudinal target SSA in 
SIDIS are
dominated by subleading-twist effects\cite{Airapetian:2005jc} and 
more difficult to interpret\cite{Mulders:1995dh,Afanasev:2003ze}.

The Collins and Sivers effects
were subject to intensive phenomenological studies in 
hadron-hadron-collisions\cite{Anselmino:1994tv}$^-$\cite{Anselmino:2004ky}
% full:
% \cite{Anselmino:1994tv,Anselmino:1998yz,D'Alesio:2004up,Anselmino:2004ky}
and in SIDIS\cite{Efremov:2001cz}$^-$\cite{Collins:2005ie}.
% full:
% \cite{Efremov:2001cz,DeSanctis:2000fh,Efremov:2003tf,Efremov:2004hz,Efremov:2004tp,Anselmino:2005nn,Vogelsang:2005cs,Collins:2005ie,in-progress}.
In this note we will concentrate on the Sivers effect and review our recent
work\cite{Efremov:2004tp,Collins:2005ie,in-progress}. 
Studies of the equally interesting Collins effect are reported 
elsewhere\cite{Vogelsang:2005cs,talk-at-SIR05}.

The Sivers function belongs to the class of so-called 
``naively time-reversal-odd'' distributions, which have been predicted to
obey an unusual ``universality property'', namely to appear with opposite 
sign in SIDIS and in the Drell-Yan (DY) process\cite{Collins:2002kn}.
We show that an experimental test of this prediction, which is among 
the most important issues for the future spin physics, is feasible in the
running or planned experiments at RHIC, COMPASS and GSI.

%===================  SECTION 2: SIVERS IN SIDIS =====================
\section{Sivers function from preliminary $P_{h\perp}$-weighted data [1]}
\label{Sec-2:prelim-Pperp-weighted}

In SIDIS the Sivers effect gives rise to a transverse target SSA with a 
specific angular distribution of the produced hadrons 
$\propto\sin(\phi-\phi_S)$, where $\phi$ ($\phi_S$) is the 
azimuthal angle of the produced hadron (target polarization vector) with 
respect to the axis defined by the hard virtual photon\cite{Boer:1997nt}.

Weighting the events entering the spin asymmetry with 
$\sin(\phi-\phi_S)P_{h\perp}$ ($P_{h\perp}=$ transverse 
momentum of the produced hadron) yields the Sivers SSA given\cite{Boer:1997nt}
(if we neglect soft factors\cite{Collins:1981uk,Ji:2004wu,Collins:2004nx}) by
\be\label{Eq:AUT-Siv-unw-SIDIS-exp}
        A_{UT}^{\sin(\phi-\phi_S)P_{h\perp}/M_N}
        = (-2) \;
        \frac{\sum_a e_a^2\,x f_{1T}^{\perp(1)a}(x)\,z D_1^{a/\pi}(z)}{
              \sum_a e_a^2\,x f_1^a(x)\,D_1^{a/\pi}(z)} \;,
\ee
where the transverse moment of the Sivers function is defined as
\be\label{Eq:Def-Siv-transverse-mom}
        f_{1T}^{\perp(1)a}(x) \equiv \int\!\di^2{\bf p}_T\;
        \frac{{\bf p}_T^2}{2 M_N^2}\;f_{1T}^{\perp a}(x,{\bf p}_T^2) \;.
\ee
{\sl Preliminary} data analyzed in this way are 
available\cite{HERMES-new}. Neglecting $f_{1T}^{\perp\bar q}$ and 
taking the ansatz motivated by predictions from the large-$N_c$ 
limit\cite{Pobylitsa:2003ty}
\be\label{Eq:large-Nc}
         f_{1T}^{\perp(1) u}(x)\;\stackrel{{\rm large}\;N_c}{=}\;
        -f_{1T}^{\perp(1) d}(x)\;\stackrel{\rm ansatz}{=}\;
        A\,x^b(1-x)^5 \;,\ee 
the Sivers function was extracted from these data\cite{Efremov:2004tp}. The 
fit result ($\chi^2$ per degree of freedom $\equiv\chi^2_{\rm dof}\sim 0.3$) 
refers to a scale of about $2.5\,{\rm GeV}^2$ and is shown in 
Fig.~\ref{Fig1-Phperp+fit}. For $f_1^a$ and $D_1^a$ the
parameterizations\cite{Gluck:1998xa,Kretzer:2001pz} were used.

\section{Transverse parton momenta and the Gaussian ansatz}

However, the currently available {\sl published} 
data\cite{Airapetian:2004tw,Alexakhin:2005iw} were analyzed without a 
$P_{h\perp}$-weight, and can only be interpreted by resorting to some 
{\sl model} for the distribution of the transverse parton momenta in the 
``unintegrated'' distribution and fragmentation functions\cite{Collins:2003fm}.
Here (for other models see\cite{Anselmino:2005nn,Vogelsang:2005cs}) 
we assume the distributions of transverse parton momenta to be Gaussian:
\ba\label{Eq:Gauss-ansatz}
        f_1^a(x,{\bf p}_T^2) & \equiv & f_1^a(x)
        \frac{\exp(-{\bf p}_T^2/p^2_{\rm unp})}{\pi \; p^2_{\rm unp}}\;,
        \nonumber\\
        f_{1T}^{\perp a}(x,{\bf p}_T^2) & \equiv & f_{1T}^{\perp a}(x)\;
        \frac{\exp(-{\bf p}_T^2/p^2_{\rm Siv})}{\pi \;p^2_{\rm Siv}} \;,
        \nonumber\\
        D_1^a(z,{\bf K}_T^2) & \equiv & D_1^a(z)
        \frac{\exp(-{\bf K}_T^2/K^2_{\! D_1})}{\pi \; K^2_{\! D_1}} \; \ea
and take the Gaussian widths to be flavour and $x$- or $z$-independent.
This model describes well the distributions of low 
(with respect to the relevant hard scale) transverse hadron momenta in 
various hard reactions\cite{D'Alesio:2004up} --- which is the case
at HERMES\cite{Airapetian:2004tw}, where
$\la P_{h\perp}\ra\sim0.4\,{\rm GeV}\ll \la Q^2\ra^{1/2}\sim 1.5\,{\rm GeV}$.

In order to test the ansatz (\ref{Eq:Gauss-ansatz}) for $f_1^a$ and $D_1^a$
in SIDIS we consider the HERMES data\cite{Airapetian:2002mf} on the average 
transverse momentum of the produced hadrons given by $\la P_{h\perp}(z)\ra=$
$\frac{\sqrt{\pi}}{2}\sqrt{z^2 p^2_{\rm unp}+K^2_{\!D_1}}$
in the Gaussian ansatz\cite{Collins:2005ie}.

With the (fitted) parameters $p^2_{\rm unp}= 0.33\,{\rm GeV}^2$ and
$K^2_{\! D_1} = 0.16\,{\rm GeV}^2$ the Gaussian ansatz provides a good
description of the data\cite{Airapetian:2002mf} --- Fig.~\ref{Fig1-Phperp+fit}.
For comparison we also show the description one obtains using the 
parameters obtained from an analysis\cite{Anselmino:2005nn} of EMC 
data\cite{Arneodo:1986cf} on the Cahn effect\cite{Cahn:1978se}, 
which is equally satisfactory. 
The good agreement observed in Fig.~\ref{Fig1-Phperp+fit} indicates that the 
mechanisms generating the Cahn effect in the EMC data\cite{Arneodo:1986cf}
and transverse hadron momenta at HERMES\cite{Airapetian:2002mf} 
could be compatible\cite{Anselmino:2005nn,Collins:2005ie}.

\section{Sivers function from final (published) HERMES data [2]}
\label{Sec-4:final-non-Pperp-weighted}

In the Gaussian model the expression for the Sivers SSA weighted without
a power of transverse hadron momentum is given by\cite{Efremov:2003tf}
\be\label{Eq:AUT-SIDIS-Gauss}
        A_{UT}^{\sin(\phi-\phi_S)} = a_{\rm Gauss}^{\rm SIDIS}\;
        A_{UT}^{\sin(\phi-\phi_S)P_{h\perp}/M_N}
        \; ; \;\;
        a_{\rm Gauss}^{\rm SIDIS} = \frac{\sqrt{\pi}}{2}\;
        \frac{M_N}{\sqrt{p^2_{\rm Siv}+K^2_{\! D_1}/z^2}}
        \;.\ee
Positivity\cite{Bacchetta:1999kz} constrains $p^2_{\rm Siv}$ to be in the 
range\cite{Collins:2005ie}
$0 < p^2_{\rm Siv}<0.33 \,{\rm GeV}^2$. Though vague, this information is
{\sl sufficient} for the extraction of the {\sl transverse moment}
of the Sivers function.

%------ BEGIN FIGURE 1: Phperp(z) at HERMES --------------------------
\begin{figure}[ht]\epsfxsize=10cm   
\centerline{
        \epsfxsize=1.4in\epsfbox{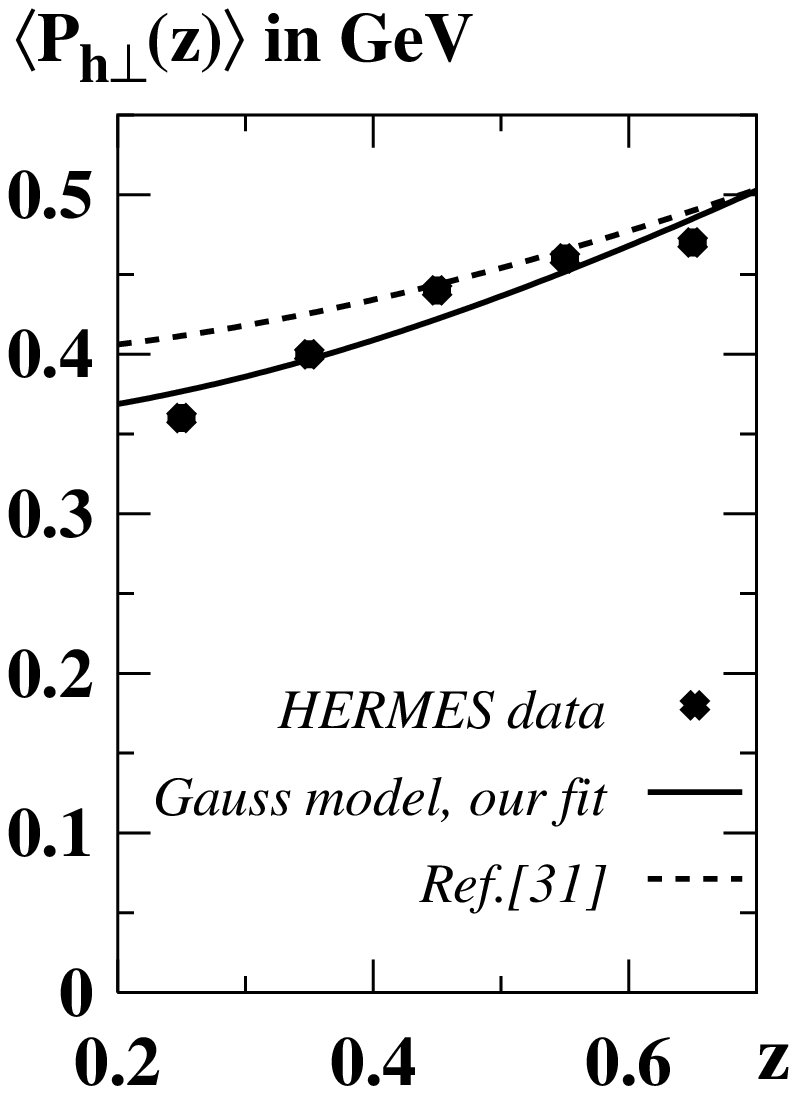}
        \hspace{-0.6cm}
        \epsfxsize=1.73in\epsfbox{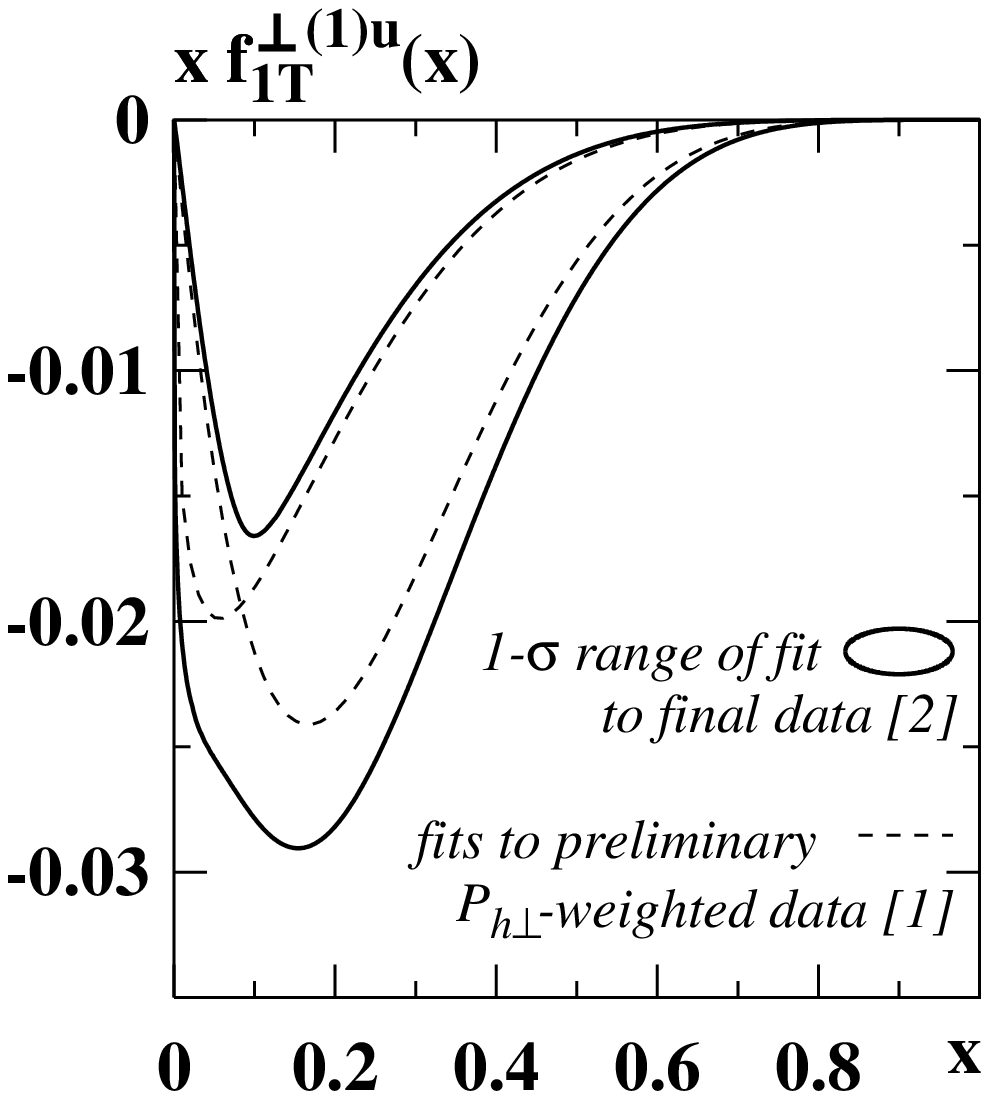}
        \hspace{-0.5cm}
        \epsfxsize=1.73in\epsfbox{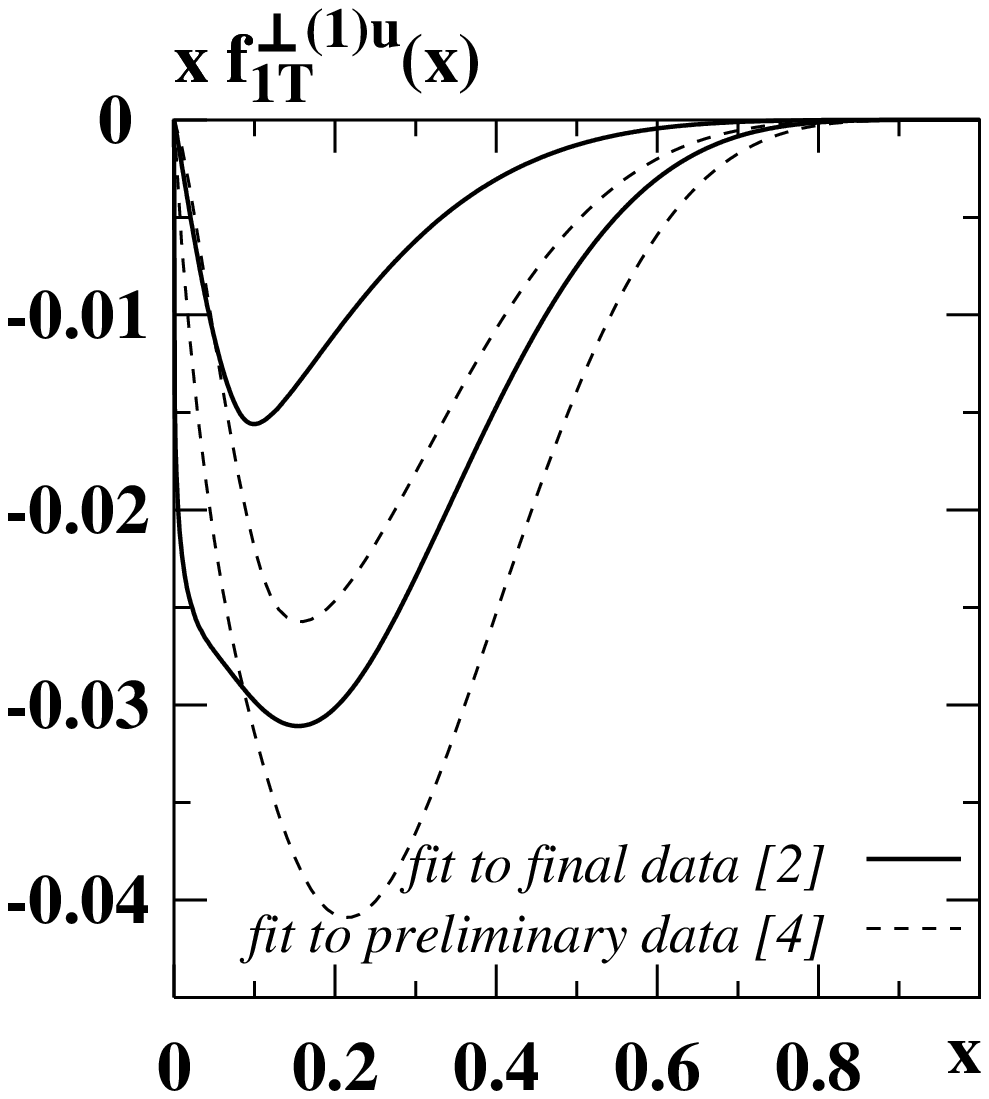}}
\caption{
        \label{Fig1-Phperp+fit}
Left:   $\la P_{h\perp}(z)\ra$ of pions produced at 
        HERMES$^{17}$ %\cite{Airapetian:2002mf} 
        vs.~$z$. The dashed (dotted) curve follows from the Gaussian ansatz with 
        the parameters from our approach$^{33}$ %\cite{Collins:2005ie}
        (from a study of the Cahn effect$^{31}$). %\cite{Anselmino:2005nn}.
Middle: The two alternative fits$^{30}$ %\cite{Efremov:2004tp} 
        of $xf_{1T}^{\perp(1)u}(x)$ to the $P_{h\perp}$-weighted 
        {\sl preliminary} data$^1$ %\cite{HERMES-new}
        and the 1-$\sigma$ region of the fit$^{33}$ %\cite{Collins:2005ie} 
        to {\sl final} (non-$P_{h\perp}$-weighted) 
        data$^2$ %\cite{Airapetian:2004tw}.
Right:  The 1-$\sigma$ regions of the fits 
        to the final data$^2$ %\cite{Airapetian:2004tw}
        and to the preliminary data$^4$.} %\cite{Diefenthaler:2005gx}
\end{figure}
%-------- END FIGURE 1. -----------------------------------------------

Using the same assumptions (large-$N_c$, neglect of $\bar q$, etc.) as in 
Sec.~\ref{Sec-2:prelim-Pperp-weighted} the fit ($\chi^2_{\rm dof}\sim 0.3$) 
to the final data\cite{Airapetian:2004tw} was obtained shown in 
Fig.~\ref{Fig1-Phperp+fit}. The fit\cite{Collins:2005ie} to the final 
data\cite{Airapetian:2004tw} obtained {\sl assuming the Gaussian model} 
is compatible --- see Fig.~\ref{Fig1-Phperp+fit} ---
with the {\sl ``model-independent''} fit\cite{Efremov:2004tp} 
to the preliminary data\cite{HERMES-new}. This observation is a
valuable test of the Gaussian ansatz for the Sivers function.

\section{Sivers effect from deuteron at COMPASS [3]}

At COMPASS the deuteron Sivers effect was found consistent with zero within 
error bars\cite{Alexakhin:2005iw}. Notice that the Sivers SSA from deuteron 
is sensitive solely to $(f_{1T}^{\perp u}+f_{1T}^{\perp d})$ which is 
subleading in the large-$N_c$ limit. 
Thus, the deuteron Sivers SSA is $\sim{\mathcal O}(N_c^{-1})$, 
while the proton Sivers SSA is $\sim{\mathcal O}(N_c^0)$.

This suppression
naturally explains\cite{Efremov:2004tp,Collins:2005ie} the compatibility of 
the HERMES and COMPASS
results\cite{Airapetian:2004tw,Alexakhin:2005iw}, within errors.
The COMPASS data\cite{Alexakhin:2005iw} confirm the utility 
of the constraint (\ref{Eq:large-Nc}) {\sl at the present stage}.

\section{Sivers function from most recent preliminary data [4]}

Increasing precision of data will, sooner or later, require to relax the 
strict large-$N_c$ constraint (\ref{Eq:large-Nc}). The {\sl preliminary} 
HERMES data released recently\cite{Diefenthaler:2005gx}
are considerably more precise compared to the final (published) 
data\cite{Airapetian:2004tw}. 

Could these data already constrain
$1/N_c$-corrections?
In order to answer this question, we repeat here 
the procedure of [\refcite{Collins:2005ie}] described 
in Sec.~\ref{Sec-4:final-non-Pperp-weighted} with the 
preliminary HERMES data\cite{Diefenthaler:2005gx}.  
The resulting fit is compatible with the fit obtained from the
published data\cite{Airapetian:2004tw}
--- Fig.~\ref{Fig1-Phperp+fit}.

The $\chi^2_{\rm dof}\sim 2$ of this fit is larger than
previously, see Sec.~\ref{Sec-4:final-non-Pperp-weighted}, which
indicates that the description of the {\sl preliminary} 
data\cite{Diefenthaler:2005gx} could be improved, e.g., by considering 
$1/N_c$ corrections. However, it could be equally sufficient to introduce 
more parameters in the large-$N_c$ ansatz (\ref{Eq:large-Nc}).
Thus, our large-$N_c$ ansatz is still useful to describe the most recent 
{\sl preliminary} data\cite{Diefenthaler:2005gx}.

% Studies\cite{Anselmino:2005nn,Vogelsang:2005cs} of the {\sl preliminary} 
% data\cite{Diefenthaler:2005gx}, where no use was made of large-$N_c$
% relations, show in fact a $|(f_{1T}^{\perp u}+f_{1T}^{\perp d})(x)|\neq 0$
% albeit substantially smaller than $|(f_{1T}^{\perp u}-f_{1T}^{\perp d})(x)|$
% in agreement with theoretical expectations\cite{Pobylitsa:2003ty}.

%===================  SECTION 3: SIVERS IN DY ========================
\section{Sivers effect in the Drell-Yan process}

On the basis of the first study\cite{Efremov:2004tp} of the preliminary 
$P_{h\perp}$-weighted data\cite{HERMES-new} it was found that the Sivers 
effect can give rise to SSA in DY large enough to be measured in the planned 
COMPASS\cite{COMPASS-proposal} and PAX\cite{PAX,PAX-estimates} experiments. 
This conclusion is now solidified\cite{Collins:2005ie,in-progress}
by the study of the published HERMES data\cite{Airapetian:2004tw}.
Thus, the predicted\cite{Collins:2002kn} sign reversal of the {\sl quark}
Sivers function in SIDIS and DY can be tested at COMPASS (PAX) in
$p^\uparrow\pi^-$ ($p^\uparrow\bar{p}$) collisions.

At RHIC in DY from $p^\uparrow p$-collisions,
however, {\sl antiquark} Sivers distributions are of importance, which are 
not constrained by the present SIDIS data\cite{Collins:2005ie}.
In order to see, what one can learn from RHIC about the Sivers SSA, 
let us assume the $\bar q$-Sivers distributions are given by
\ba
\label{Eq:model-Sivers-qbar} 
        f_{1T}^{\perp(1) \bar q}(x) = 
        f_{1T}^{\perp(1)      q}(x) \times 
        \left\{
        \begin{array}{ll}
 0.25 = {\rm const}                                     \;\;&\mbox{(model I)}\\
 \frac{(f_1^{\bar u}+f_1^{\bar d})(x)}{(f_1^u+f_1^d)(x)}\;\;&\mbox{(model II),}
        \end{array} 
        \right.  
\ea
with $f_{1T}^{\perp(1) q}(x)$ from the fit to the published HERMES 
data\cite{Airapetian:2004tw} --- Sec.~\ref{Sec-4:final-non-Pperp-weighted}.
The {\sl models} I, II are consistent\cite{Collins:2005ie} 
with theoretical constraints, {\sl and} with SIDIS 
data\cite{HERMES-new,Airapetian:2004tw,Alexakhin:2005iw}.
This makes them well suited to visualize possible effects of 
$\bar q$-Sivers distributions at RHIC.

The Sivers SSA in DY is defined similarly to that in SIDIS. 
Here $(\phi-\phi_S)$ is the azimuthal angle between the virtual photon and 
the polarization vector (the polarized proton moves into the positive
$z$-direction). We again neglect soft factors and assume the Gaussian
model, so that the SSA is
\be\label{Eq:DY-AUT-1}
        A_{UT}^{\sin(\phi-\phi_S)} = 2\;\frac{a_{\rm Gauss}^{\rm DY}
\sum_a e_a^2\,x_1f_{1T\,\,{\rm DY}}^{\perp(1) a}(x_1)\,x_2f_1^{\bar a}(x_2)}{
\sum_a e_a^2\,x_1f_1^{a}                        (x_1)\,x_2f_1^{\bar a}(x_2)}
        \,,\ee
where $x_{1,2}=(Q^2/s)^{1/2}\,e^{\pm y}$ with $s=(p_1+p_2)^2$, 
$Q^2=(k_1+k_2)^2$, and $y=\frac12\,{\rm ln}\frac{p_1\cdot(k_1+k_2)}{p_2\cdot(k_1+k_2)}$.
The momenta of the incoming proton (outgoing lepton) pair are denoted by
$p_{1/2}$ ($k_{1/2}$). The Gaussian factor reads
\be\label{Eq:DY-AUT-1a}
        a_{\rm Gauss}^{\rm DY} = \frac{\sqrt{\pi}}{2}\,
        \frac{M_N}{\sqrt{p^2_{\rm Siv} 
                        +p^2_{\rm unp}\!}\,}\;.
\ee

\begin{figure}[ht]\epsfxsize=10cm   
\vspace{-0.5cm}

\centerline{\epsfxsize=2.3in\epsfbox{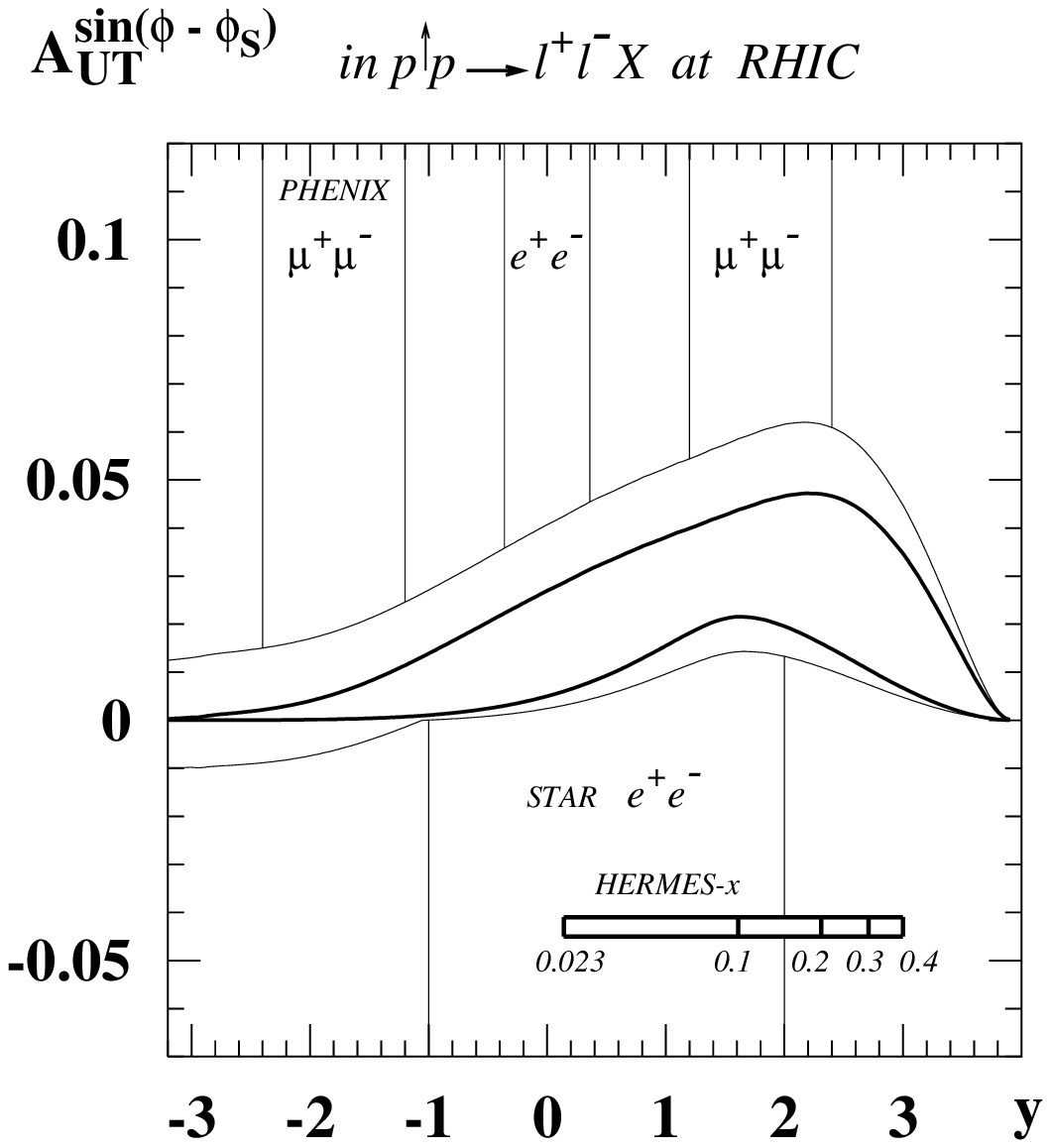}\hspace{-0.4cm}
            \epsfxsize=2.3in\epsfbox{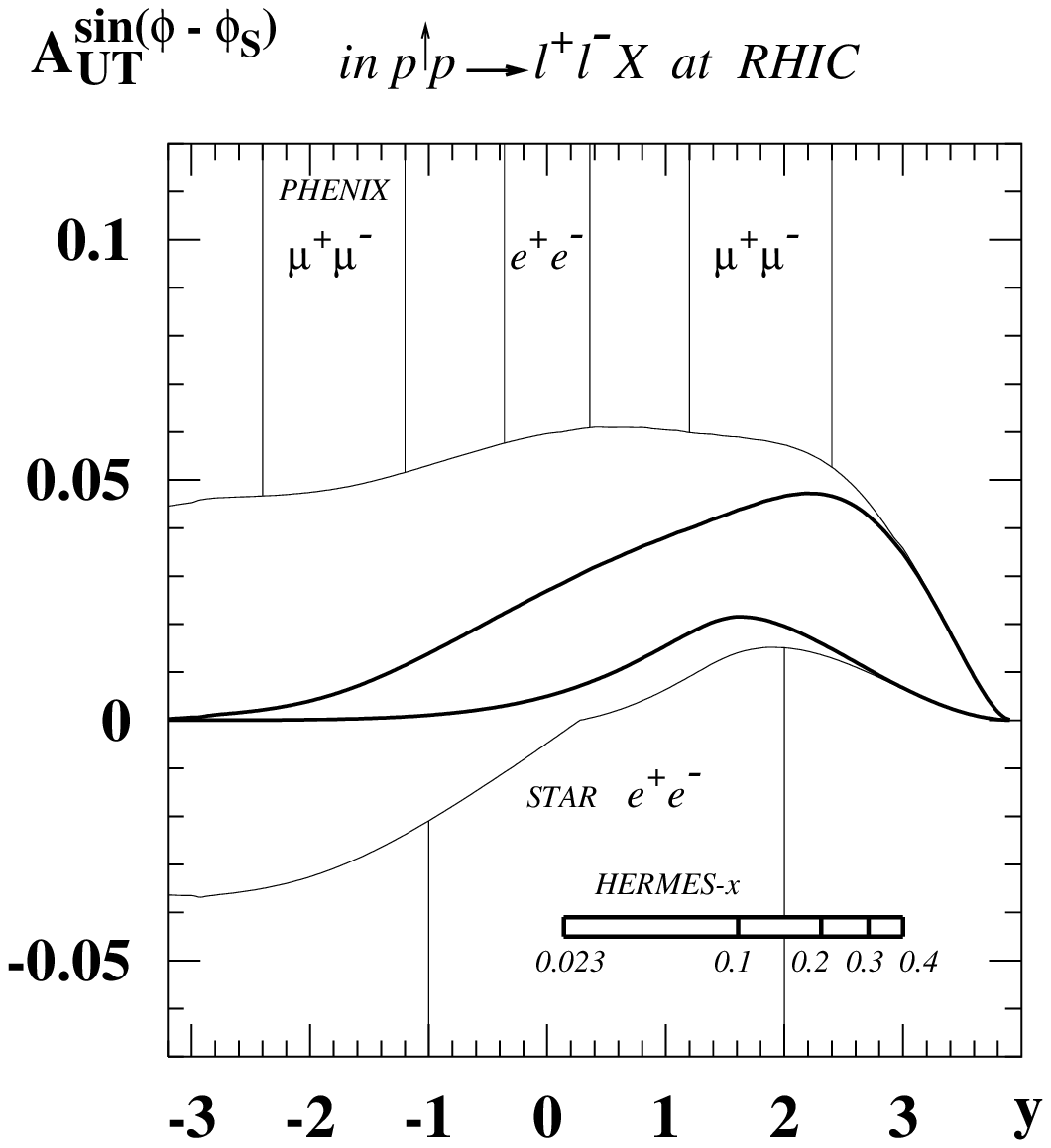}}
\caption{
    \label{Fig5-DY-at-RHIC-new} 
    The Sivers SSA $A_{UT}^{\sin(\phi-\phi_S)}$ in $p^\uparrow p\to l^+l^- X$
    as function of $y$ for the kinematics of the RHIC experiment
    with $\sqrt{s}=200\,{\rm GeV}$, and $Q^2=(4\,{\rm GeV})^2$.
    The inner error band (thick lines) shows the 1-$\sigma$ uncertainty
    of the fit$^{33}$. %\cite{Collins:2005ie}. 
    The $x$-region explored at HERMES is included to indicate where
    the SIDIS data constrain the prediction.
    The outer error band (thin lines) arises from assuming Sivers 
    $\bar q$-distribution functions according to model I (left)
    and model II (right) in Eq.~(\ref{Eq:model-Sivers-qbar}). 
    Also shown are the regions in which PHENIX and STAR can detect 
    $\mu^+\mu^-$ or $e^+e^-$ pairs.}
\end{figure}

Considering the sign reversal\cite{Collins:2002kn} we obtain the results 
shown in Fig.~\ref{Fig5-DY-at-RHIC-new}. In our estimate we assume the ratio
$\sum_ae_a^2f_{1T}^{\perp(1)a}f_1^{\bar a}/\sum_b e_b^2f_1^bf_1^{\bar b}$
to be weakly scale-dependent, and roughly simulate Sudakov dilution by 
assuming that $p^2_{\rm unp/Siv}$ increase by a factor of two from HERMES to RHIC energies. 
Notice that SSA weighted appropriately with the transverse dilepton 
momentum\cite{Efremov:2004tp} were argued to be less sensitive to Sudakov 
suppression\cite{Boer:2001he}.

In the region of positive rapidities $1\leq y\leq 2$ the Sivers SSA at RHIC 
is well constrained by the SIDIS data\cite{Airapetian:2004tw} and shows
little sensitivity to the unknown $\bar{q}$-Sivers distributions. 
Thus, in this region STAR and PHENIX can also test the sign 
reversal\cite{Collins:2002kn} of the {\sl quark} Sivers function.

For negative $y$ the Sivers SSA is strongly sensitive to the antiquark Sivers 
function --- with the effect being more pronounced at larger dilepton masses 
$Q$.\cite{Collins:2005ie}
This reveals the unique feature of RHIC, which --- in contrast to COMPASS
and PAX --- can also provide information on the antiquark Sivers distribution.

%------ BEGIN FIGURE 2: DY at RHIC ----------------------------------
%-------- END FIGURE 2. -----------------------------------------------

\section{Conclusions}

We reviewed our studies\cite{Efremov:2004tp,Collins:2005ie} 
of the HERMES and COMPASS data on the Sivers effect in 
SIDIS\cite{HERMES-new,Airapetian:2004tw,Alexakhin:2005iw,Diefenthaler:2005gx}.
The data from various targets are compatible with each other
and can be well described assuming a Gaussian distribution of parton 
transverse momenta in the distribution and fragmentation functions.
The parameters in the Gaussian ansatz were constrained by 
HERMES data and are compatible with results obtained from studies
of the Cahn effect\cite{Anselmino:2005nn}.

The 
data\cite{HERMES-new,Airapetian:2004tw,Alexakhin:2005iw,Diefenthaler:2005gx}
confirm the predictions from the large-$N_c$ limit on the flavour dependence
of the Sivers function\cite{Pobylitsa:2003ty}. The sign of the extracted
quark Sivers functions is in agreement with the intuitive picture
discussed in [\refcite{Burkardt:2002ks}].
Results by other groups\cite{Anselmino:2005nn,Vogelsang:2005cs} confirm these 
findings --- see the detailed comparison in Ref.~[\refcite{Anselmino:2005an}].

The information on the quark Sivers distributions extracted from SIDIS is 
required for reliable estimates of the Sivers effect in DY for current or 
planned experiments.
We estimated that the Sivers SSA in DY are sizeable enough to be observed
at RHIC, COMPASS and PAX\cite{Efremov:2004tp,Collins:2005ie,in-progress}
allowing one to test the QCD prediction\cite{Collins:2002kn} that the Sivers 
function should appear with opposite signs in SIDIS and in DY. 
In addition, RHIC can provide information on the antiquark Sivers 
distributions\cite{in-progress}.

\section*{Acknowledgments.}
The work 
% is partially supported by BMBF and DFG of Germany, the
% COSY-J{\"u}lich project, the Transregio Bonn-Bochum-Giessen, and 
is part of the European Integrated Infrastructure Initiative Hadron
Physics project under contract number RII3-CT-2004-506078. 
A.E. is supported by grants RFBR 03-02-16816 and DFG-RFBR 03-02-04022.
J.C.C. is supported in part by the U.S. D.O.E., and by a Mercator
Professorship of DFG.

\end{document}